\documentclass[%
reprint,
 superscriptaddress,
showpacs,
 amsmath,amssymb,
 aps,
]{revtex4-1}

\usepackage{graphicx}% Include figure files
\usepackage{dcolumn}% Align table columns on decimal point
\usepackage{bm}% bold math
\begin{document}

\preprint{APS/123-QED}

\title{Observation of Spin-Dependent Charge Symmetry Breaking 
in $\Lambda N$ Interaction:\\
Gamma-Ray Spectroscopy of $^4_{\Lambda }$He}% Force line breaks with \\

\author{T.~O.~Yamamoto}
\affiliation{Department~of~Physics,~Tohoku~University,~Sendai~980-8578,~Japan }

\author{M.~Agnello}
\affiliation{ Dipartimento di Scienza Applicate e Tecnologica, Politecnico di Torino, Corso Duca degli Abruzzi, 10129 Torino, Italy }
\affiliation{ INFN, Sezione di Torino, via P. Giuria 1, 10125 Torino, Italy}

\author{Y.~Akazawa}
\affiliation{Department~of~Physics,~Tohoku~University,~Sendai~980-8578,~Japan }

\author{N.~Amano}
\affiliation{Department of Physics, Kyoto University, Kyoto 606-8502, Japan }

\author{K.~Aoki}
\affiliation{ Institute of Particle and Nuclear Studies (IPNS), High Energy Accelerator Research Organization
(KEK), Tsukuba, 305-0801, Japan}

\author{E.~Botta}
\affiliation{ INFN, Sezione di Torino, via P. Giuria 1, 10125 Torino, Italy}
\affiliation{ Dipartimento di Fisica, Universit di Torino, Via P. Giuria 1, 10125 Torino, Italy}

\author{N.~Chiga}
\affiliation{Department~of~Physics,~Tohoku~University,~Sendai~980-8578,~Japan }

\author{H.~Ekawa}
\affiliation{ Department of Physics, Kyoto University, Kyoto 606-8502, Japan}

\author{P.~Evtoukhovitch}
\affiliation{ Joint Institute for Nuclear Research, Dubna ,Moscow Region 141980, Russia}

\author{A.~Feliciello}
\affiliation{ INFN, Sezione di Torino, via P. Giuria 1, 10125 Torino, Italy}

\author{M.~Fujita}
\affiliation{Department~of~Physics,~Tohoku~University,~Sendai~980-8578,~Japan }

\author{T.~Gogami}
\affiliation{ Department of Physics, Kyoto University, Kyoto 606-8502, Japan}

\author{S.~Hasegawa}
\affiliation{ Advanced Science Research Center (ASRC), Japan Atomic Agency (JAEA), Tokai, Ibaraki
319-1195, Japan}

\author{S.~H.~Hayakawa}
\affiliation{ Department of Physics, Osaka University, Toyonaka 560-0043, Japan}

\author{T.~Hayakawa}
\affiliation{ Department of Physics, Osaka University, Toyonaka 560-0043, Japan}

\author{R.~Honda}
\affiliation{ Department of Physics, Osaka University, Toyonaka 560-0043, Japan}

\author{K.~Hosomi}
\affiliation{ Advanced Science Research Center (ASRC), Japan Atomic Agency (JAEA), Tokai, Ibaraki
319-1195, Japan}

\author{S~.H.~Hwang}
\affiliation{ Advanced Science Research Center (ASRC), Japan Atomic Agency (JAEA), Tokai, Ibaraki
319-1195, Japan}

\author{N.~Ichige}
\affiliation{Department~of~Physics,~Tohoku~University,~Sendai~980-8578,~Japan }

\author{Y.~Ichikawa}
\affiliation{ Advanced Science Research Center (ASRC), Japan Atomic Agency (JAEA), Tokai, Ibaraki
319-1195, Japan}

\author{M.~Ikeda}
\affiliation{Department~of~Physics,~Tohoku~University,~Sendai~980-8578,~Japan }

\author{K.~Imai}
\affiliation{ Advanced Science Research Center (ASRC), Japan Atomic Agency (JAEA), Tokai, Ibaraki
319-1195, Japan}

\author{S.~Ishimoto}
\affiliation{ Institute of Particle and Nuclear Studies (IPNS), High Energy Accelerator Research Organization
(KEK), Tsukuba, 305-0801, Japan}

\author{S.~Kanatsuki}
\affiliation{ Department of Physics, Kyoto University, Kyoto 606-8502, Japan}

\author{M.~H.~Kim}
\affiliation{Departiment of Physics, Korea University, Seoul 136-713, Korea}

\author{S.~H.~Kim}
\affiliation{Departiment of Physics, Korea University, Seoul 136-713, Korea}

\author{S.~Kinbara}
\affiliation{Faculty of Education, Gifu University, Gifu 501-1193, Japan}

\author{T.~Koike}
\affiliation{Department~of~Physics,~Tohoku~University,~Sendai~980-8578,~Japan }

\author{J~.Y.~Lee}
\affiliation{ Department of Physics and Astronomy, Seoul National University, Seoul 151-747, Korea}

\author{S.~Marcello}
\affiliation{ INFN, Sezione di Torino, via P. Giuria 1, 10125 Torino, Italy}
\affiliation{ Dipartimento di Fisica, Universit di Torino, Via P. Giuria 1, 10125 Torino, Italy}

\author{K.~Miwa}
\affiliation{Department~of~Physics,~Tohoku~University,~Sendai~980-8578,~Japan }

\author{T.~Moon}
\affiliation{ Department of Physics and Astronomy, Seoul National University, Seoul 151-747, Korea}

\author{T.~Nagae}
\affiliation{ Department of Physics, Kyoto University, Kyoto 606-8502, Japan}

\author{S.~Nagao}
\affiliation{Department~of~Physics,~Tohoku~University,~Sendai~980-8578,~Japan }

\author{Y.~Nakada}
\affiliation{ Department of Physics, Osaka University, Toyonaka 560-0043, Japan}

\author{M.~Nakagawa}
\affiliation{ Department of Physics, Osaka University, Toyonaka 560-0043, Japan}

\author{Y.~Ogura}
\affiliation{Department~of~Physics,~Tohoku~University,~Sendai~980-8578,~Japan }

\author{A.~Sakaguchi}
\affiliation{ Department of Physics, Osaka University, Toyonaka 560-0043, Japan}

\author{H.~Sako}
\affiliation{ Advanced Science Research Center (ASRC), Japan Atomic Agency (JAEA), Tokai, Ibaraki
319-1195, Japan}

\author{Y.~Sasaki}
\affiliation{Department~of~Physics,~Tohoku~University,~Sendai~980-8578,~Japan }

\author{S.~Sato}
\affiliation{ Advanced Science Research Center (ASRC), Japan Atomic Agency (JAEA), Tokai, Ibaraki
319-1195, Japan}

\author{T.~Shiozaki}
\affiliation{Department~of~Physics,~Tohoku~University,~Sendai~980-8578,~Japan }

\author{K.~Shirotori}
\affiliation{ Research Center of Nuclear Physics, Osaka University, Ibaraki 567-0047, Japan}

\author{H.~Sugimura}
\affiliation{ Advanced Science Research Center (ASRC), Japan Atomic Agency (JAEA), Tokai, Ibaraki
319-1195, Japan}

\author{S.~Suto}
\affiliation{Department~of~Physics,~Tohoku~University,~Sendai~980-8578,~Japan }

\author{S.~Suzuki}
\affiliation{ Institute of Particle and Nuclear Studies (IPNS), High Energy Accelerator Research Organization
(KEK), Tsukuba, 305-0801, Japan}

\author{T.~Takahashi}
\affiliation{ Institute of Particle and Nuclear Studies (IPNS), High Energy Accelerator Research Organization
(KEK), Tsukuba, 305-0801, Japan}

\author{H.~Tamura}
\affiliation{Department~of~Physics,~Tohoku~University,~Sendai~980-8578,~Japan }

\author{K.~Tanabe}
\affiliation{Department~of~Physics,~Tohoku~University,~Sendai~980-8578,~Japan }

\author{K.~Tanida}
\affiliation{ Advanced Science Research Center (ASRC), Japan Atomic Agency (JAEA), Tokai, Ibaraki
319-1195, Japan}

\author{Z.~Tsamalaidze}
\affiliation{ Joint Institute for Nuclear Research, Dubna ,Moscow Region 141980, Russia}

\author{M.~Ukai}
\affiliation{Department~of~Physics,~Tohoku~University,~Sendai~980-8578,~Japan }

\author{Y.~Yamamoto}
\affiliation{Department~of~Physics,~Tohoku~University,~Sendai~980-8578,~Japan }

\author{S.~B.~Yang}
\affiliation{ Department of Physics and Astronomy, Seoul National University, Seoul 151-747, Korea}

\collaboration{J-PARC E13-1\textsuperscript{st} Collaboration }%\noaffiliation

\date{\today}% It is always \today, today,

\begin{abstract}
The energy spacing between the ground-state spin doublet of
$^4_\Lambda $He(1$^+$,0$^+$) was determined to be 
$1406 \pm 2 \pm 2$ keV,
by measuring $\gamma$ rays for the $1^+ \to 0^+$ transition 
with a high efficiency germanium detector array 
in coincidence with the $^4$He$(K^-,\pi^-)$ $^4_\Lambda $He reaction 
at J-PARC.
In comparison to the corresponding energy spacing 
in the mirror hypernucleus $^4_\Lambda $H, 
the present result clearly indicates the existence of 
charge symmetry breaking (CSB) in $\Lambda N$ interaction.
It is also found that the CSB effect is large in the $0^+$ ground state
but is by one order of magnitude smaller in the $1^+$ excited state,
demonstrating that the $\Lambda N$ CSB interaction has 
spin dependence.

% \begin{description}
% Secondary publications and information retrieval purposes.
% \item[PACS numbers]
% May be entered using the \verb+\pacs{#1}+ command.
% You may use the \texttt{description} environment to structure your abstract;
% use the optional argument of the \verb+\item+ command to give the category of each item. 
% \end{description}
\end{abstract}


\maketitle

%%------------------------------------
%%------------------------------------
%%------------------------------------

Charge symmetry is a basic symmetry in nuclear physics which governs properties and structures of atomic nuclei.
%Since charge symmetry comes from isospin invariance,
It should also hold in $\Lambda$$N$ interaction
and $\Lambda$ hypernuclei; 
$\Lambda$$p$ and $\Lambda$$n$ interactions and the $\Lambda$ binding energies ($B_\Lambda$)
between a pair of mirror $\Lambda$ hypernuclei such as $^4_\Lambda$H and $^4_\Lambda$He
are expected to be identical under this symmetry.

In $NN$ interaction and ordinary nuclei,
effects of charge symmetry breaking (CSB) have been observed, 
for example, in the $^3$H and $^3$He mass difference of 70 keV
and the $nn$ and $pp$ scattering length difference of $a_{nn} -a_{pp} = -1.5 \pm 0.5$ fm
(both corrected for large Coulomb effects).
In meson-exchange models,
those effects are explained by $\rho^0-\omega$ mixing (see Ref.~\cite{MILLER}, for example).

On the other hand, there has been a long standing puzzle
in CSB for $\Lambda N$ interaction. 
Old experiments using emulsion technique reported $B_\Lambda $ 
of the ground states of
$^4_\Lambda $H($0^+$) and $^4_\Lambda $He($0^+$)
to be $2.04\pm 0.04$ MeV and $2.39\pm 0.03$ MeV, respectively \cite{juric},
giving a $B_\Lambda$ difference 
$\Delta B_\Lambda(0^+)$ = 
$B_\Lambda$($^4_\Lambda$He($0^+$)) $-$ $B_\Lambda$($^4_\Lambda$H($0^+$))
= $0.35 \pm 0.05$ MeV.
Theoretical efforts have been made since 1960s \cite{DALITZ} to explain the $\Delta B_\Lambda(0^+)$ value,
 but any quantitative studies fail to give a $\Delta B_\Lambda(0^+)$ value larger than 100 keV; for example, 
a 4-body $YNNN$ coupled-channel calculation with $Y$ = $\Lambda$ and $\Sigma$ 
using the widely-accepted baryon-baryon interaction model (NSC97e) 
gives $\Delta B_\Lambda (0^+) \sim 70$ keV \cite{NOGGA}.

To approach this problem, confirmation and improvement of experimental data on CSB are also necessary.
Systematic errors are not shown in the old emulsion data for $B_\Lambda$, and thus new data,
hopefully by different experimental methods, have been awaited.
Recently, the $\pi^-$ momentum in the $^4_\Lambda$H $\to$ $^4$He $+$ $\pi^-$ weak decay
was precisely measured at MAMI-C \cite{MAMI},
and the obtained value of $B_\Lambda$($^4_\Lambda$H($0^+$)) = 2.12 $\pm$ 0.01 (stat.) $\pm$  0.09 (syst.) MeV
is consistent with the emulsion value.

\begin{figure}[t]    
\begin{center}
\includegraphics[width=8.7cm]{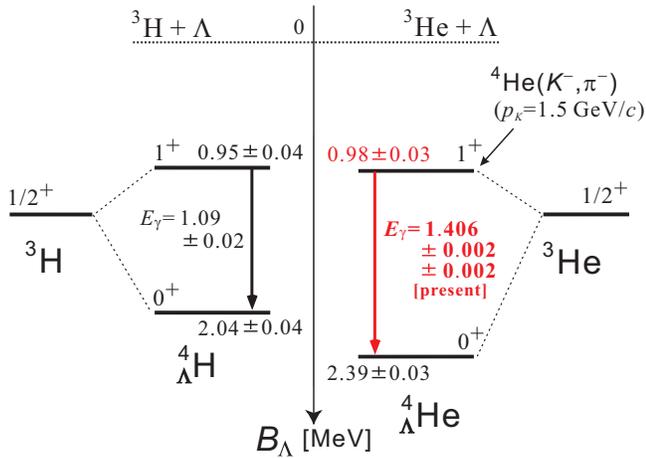}
\caption{
(color online).
Level schemes of the mirror hypernuclei,
$^4_\Lambda $H and $^4_\Lambda $He.
$\Lambda $ binding energies ($B_\Lambda $) of $^4_\Lambda$H($0^+$)
and $^4_\Lambda$He($0^+$) 
are taken from past emulsion experiments \cite{juric}.
$B_\Lambda $($^4_\Lambda$He($1^+$)) and $B_\Lambda $($^4_\Lambda$H($1^+$))
are obtained using the present data and 
past $\gamma$-ray data \cite{H_1,old_He,Kawachi}, respectively.
Recently,
$B_\Lambda$($^4_\Lambda$H($0^+$)) = 2.12 $\pm$ 0.01 (stat.) $\pm$  0.09 (syst.) MeV
was obtained with an independent technique \cite{MAMI}.
% The $1^+$ state of $^4_\Lambda $He was populated
% by the 1.5 GeV/$c$ $(K^-,\pi^-)$ reaction in the present work.
% The measured excitation energies are also given.
}
\label{He_level}
\end{center}
\end{figure}

In addition, the $B_\Lambda$ difference for the excited $1^+$ states provides an important information
on the spin dependent CSB effect from which the origin of CSB can be studied. 
The $B_\Lambda$ values for the $1^+$ state are obtained via the $1^+ \to 0^+$ $\gamma$-ray transition energies.
The $^4_{\Lambda}$H $\gamma$ ray was measured three times, 
and the $^4_{\Lambda}$H($1^+,0^+$) energy spacing was determined
to be 1.09 $\pm$ 0.02 MeV as the wighted average of these three
(1.09 $\pm$ 0.03 MeV \cite{H_1},
1.04 $\pm$ 0.04 MeV \cite{old_He},  and
1.114 $\pm$ 0.030 MeV \cite{Kawachi}),
as shown in Fig.~\ref{He_level} (left).
On the other hand,
observation of the $^4_{\Lambda}$He $\gamma$ ray was reported only once
by an experiment with stopped $K^-$ absorption on a $^7$Li target,
which claimed the ($1^+,0^+$) energy spacing of 
1.15 $\pm$ 0.04  MeV \cite{old_He}.
This result suggests a significantly large CSB effect also
in the $1^+$ state with $\Delta B_\Lambda (1^+) $ = 0.29 $\pm$ 0.06 MeV.
However, this $^4_\Lambda$He $\gamma$-ray spectrum is statistically 
insufficient, and identification of $^4_\Lambda$He hyperfragment
through high energy $\gamma$ rays attributed
to the $^4_\Lambda$He $\to$ $^4$He + $\pi^0$ weak decay seems to be ambiguous.

In order to clarify this situation,
we performed a $\gamma$-ray spectroscopy experiment of $^4_{\Lambda}$He at J-PARC \cite{TamuraA},
in which the $1^+$ excited state of $^4_\Lambda$He was directly produced
via the $^4$He$(K^-,\pi^-)$ reaction
with a 1.5 GeV/c $K^-$ beam,  and $\gamma$ rays were measured using germanium (Ge) detectors
with an energy resolution by one order of magnitude better than
NaI counters used in all the previous $^4_\Lambda$H and $^4_\Lambda$He $\gamma$-ray experiments.
In this letter, we present the result which clearly supersedes the
previously claimed $\gamma$-ray transition energy and firmly 
establishes the level scheme of $^4_\Lambda$He as shown in Fig.~\ref{He_level} (right).

% ** experiment ** (this title will be removed) \\

\begin{figure}[t]    
\begin{center}
\includegraphics[width=8.5cm]{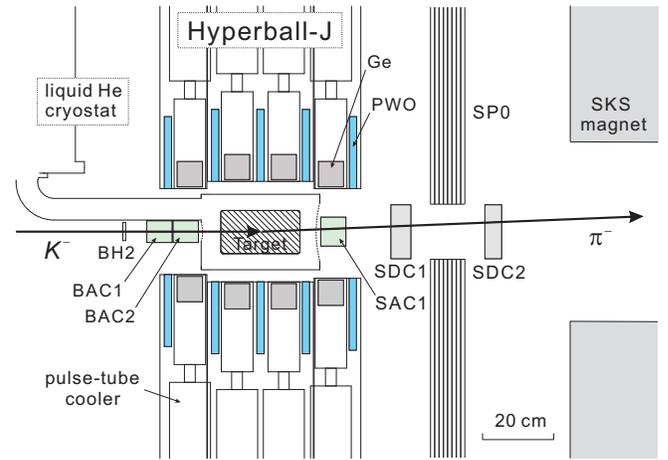}
\caption{
(color online).
A schematic view of experimental setup around the liquid $^4$He target (side view).
SKS is a superconducting dipole magnet;
BH2 is a plastic scintillation counter hodoscope;
BAC1,2 and SAC1 are aerogel $\check{\rm C}$erenkov counters with n = 1.03;
SDC1,2 are drift chambers.
SP0 is an electromagnetic shower counter to tag high energy photons from $\pi^0$ decay.  
Hyperball-J consists of 27 Ge detectors, each surrounded by PWO counters for background suppression.
}
\label{HBJ}
\end{center}
\end{figure}

The J-PARC E13-1\textsuperscript{st} experiment was carried out at
the K1.8 beam line in the J-PARC Hadron Experimental Facility \cite{PTEP_agari}.
The $^4$He$(K^-,\pi^-)$ reaction was used to produce $^4_{\Lambda }$He($1^+$),
which was populated via the spin-flip amplitude of the $K^-$ + $n$ $\to$ $\Lambda$ + $\pi^-$ process.
A beam momentum of 1.5 GeV/$c$ was chosen considering the elementary cross section of
the spin-flip $\Lambda $ production and the available beam intensity.
A 2.7 g/cm$^2$-thick liquid $^4$He target was irradiated with
a total of $2.3\times 10^{10}$ kaons.
A high purity $K^-$ beam with a $K^-/\pi^-$ ratio of $\sim $2
was delivered to the target with a typical intensity of $3\times 10^5$ over a 
2.1 s duration of the beam spill occurring every 6 s.
Incident $K^-$ and outgoing $\pi^-$ mesons
were particle-identified and momentum-analyzed
by the beam line spectrometer
and the Superconducting Kaon Spectrometer (SKS) \cite{PTEP_takahashi}, respectively.
On the other hand, $\gamma $ rays are detected by a Ge detector array (Hyperball-J) surrounding the target.
Through  coincidence measurement between these spectrometer systems
and Hyperball-J, $\gamma $ rays from hypernuclei were measured.
The detector system surrounding the target is shown in Fig.~\ref{HBJ}.

In SKS, detector setting was configured
for $\gamma$-ray spectroscopy experiments via the $(K^-, \pi^-)$ reaction (SksMinus).
SksMinus had a large acceptance for detecting the outgoing pions
in the laboratory scattering angle range of $\theta _{K\pi}$= 0$^\circ$--20$^\circ$.
The $(K^-,\pi^-)$ reaction events were
identified with threshold-type aerogel $\check{\rm C}$erenkov counters
at the trigger level and by time of flight in the off-line analysis.
The $^4_\Lambda $He mass was calculated as the missing mass
of the $^4$He$(K^-,\pi^-)$ reaction.
Detailed description of the spectrometer system and of the analysis procedure for
calculating missing mass will be reported elsewhere.

Hyperball-J is a newly developed Ge detector array for  hypernuclear $\gamma $-ray spectroscopy \cite{HBJ}.
The array can be used in a high intensity hadron beam condition
by introducing mechanical cooling of  Ge detectors \cite{PTR}.
The array consisted of 27 Ge detectors in total,
equipped with PWO counters surrounding each Ge crystal
to suppress background events
such as Compton scattering and high energy photons from $\pi ^0$ decay.
The Ge detectors were of coaxial type with a 60\% relative efficiency.
The Ge crystals covered a solid angle of 0.24$\times 4\pi $ sr
in total with the source point at the center.
The total absolute photo-peak efficiency was $\sim $4\% for 1-MeV $\gamma $ rays
when taking account of self absorption in the target material. 
Energy calibration was performed over the 0.6--2.6 MeV range,
by using data taken in the period off the beam spill
with thorium-series $\gamma $ rays. 
The systematic error in the energy calibration
was estimated to be 0.5 keV for that energy region.
The energy resolution was 5 keV (FWHM) at 1.4 MeV after summing up data for all the detectors.
The resolution was  slightly worse in the period on the beam spill.

\begin{figure}[t]    
\begin{center}
\includegraphics[width=8.5cm]{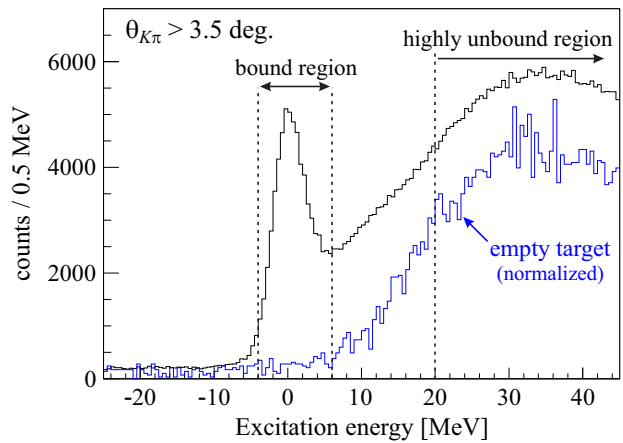}
\caption{
(color online).
The missing mass spectrum for the $^4$He$(K^-,\pi^-)^4_\Lambda $He kinematics
plotted as a function of the excitation energy, $E_{\rm ex}$,
where events with scattering angles ($\theta _{K\pi}$) larger than 3.5$^\circ$ are selected.
Black and blue lines show a spectrum with and without liquid helium, respectively.
}
\label{He_MissMass}
\end{center}
\end{figure}

\begin{figure}[t]    
\begin{center}
\includegraphics[width=8.5cm,angle=0]{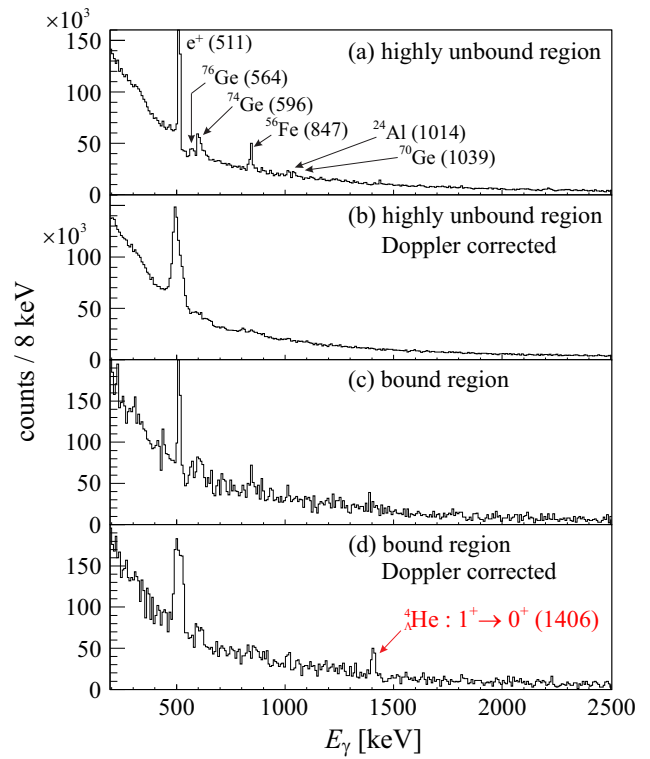}
\caption{
(color online).
$\gamma $-ray energy spectra measured by Hyperball-J
in coincidence with the $^4$He$(K^-,\pi^-)$ reaction.
Missing mass selections are applied to the highly unbound region ($E_{\rm ex} >$ +20 MeV) for (a) and (b),
and to $^4_\Lambda $He bound region ($-4 < E_{\rm ex} < +6 $ MeV) for (c) and (d).
An event-by-event Doppler correction is applied for (b) and (d). 
Single peak is observed in (d) attributed to the $M1(1^+\rightarrow 0^+)$ transition. 
}
\label{He_Gamma}
\end{center}
\end{figure}

Selected events were those in which a Ge detector has a hit
in a typical time gate of 50 ns and
without any hits in the corresponding PWO counters in
the 50 ns coincidence gate.
In the $(K^-,\pi^-)$ reaction at 1.5 GeV/$c$, produced hypernuclei have recoil velocities of
ƒÀ = 0.03--0.10, which lead to a stopping time longer than 20 ps in the target material.
The $^4_\Lambda$He($1^+ \to 0^+$) $M1$ transition with an energy of  $\sim $1 MeV is estimated to 
have a lifetime of $\sim$ 0.1 ps
assuming weak coupling between the core nucleus and the $\Lambda $ \cite{Dalitz_and_Gal}.
Therefore, the $\gamma$-ray peak shape is expected to be Doppler broadened.
We applied an event-by-event correction to the $\gamma $-ray energy
by using the measured recoil momentum of $^4_\Lambda $He,
the reaction vertex position, and the position of the Ge detector.
%This method has been employed in the previous
%hypernuclear $\gamma $-ray spectroscopy experiments \cite{Ukai_18O}. 
It is noted that the Doppler shift
correction leaves 0.1\% uncertainty in the
measured $\gamma $-ray energy,
where the dominant contribution comes from
uncertainties ($\pm $5 mm) associated with positions of
the Hyperball-J apparatus
with respect to the magnetic spectrometer systems.
Details of the analysis procedures are almost the same as the previous 
hypernuclear 
$\gamma$-ray spectroscopy experiments \cite{Ukai_18O}.

Figure~\ref{He_MissMass} shows the missing mass spectrum for $^4_\Lambda $He
as a function of the excitation energy, $E_{\rm ex}$.
Events with scattering angles ($\theta _{K\pi}$) larger than 3.5$^\circ$ were selected
to reduce the background due to beam $K^-\rightarrow \pi^- + \pi^0 $ events
which kinematically overlap with hypernuclear production events at $\theta _{K\pi}$= 0$^\circ$--3$^\circ$.
The background spectrum associated with materials other than liquid helium
as well as with $K^-$ beam decay events was obtained
with the empty target vessel as shown together in Fig.~\ref{He_MissMass};
it is evident that
the observed peak is originated from
the $^4$He$(K^-, \pi^-)$ reaction.
According to a theoretical calculation,
the $^4_{\Lambda}$He$(0^+)$ ground state is predicted to be predominantly populated,  
while the $^4_{\Lambda}$He$(1^+)$ excited state
at a lower rate ($\sim$ 1/4 of $^4_{\Lambda}$He($0^+$)) \cite{Harada_private}.
Therefore, the obtained peak is composed of $^4_{\Lambda}$He$(0^+)$ with a small contribution from  
$^4_{\Lambda}$He$(1^+)$.
The peak width of 5 MeV (FWHM) corresponds to the missing mass resolution.
The energy region for bound $^4_\Lambda$He is $E_{\rm ex} $ = 0 -- 2.39 MeV (see Fig.~\ref{He_level}). 
Thus, the region of $-4 < E_{\rm ex} < +6 $ MeV was chosen for event selection of
the $^4_\Lambda $He bound state that is allowed for  $\gamma $ decay.

\begin{figure}[t]    
\begin{center}
\includegraphics[width=8.0cm,angle=0]{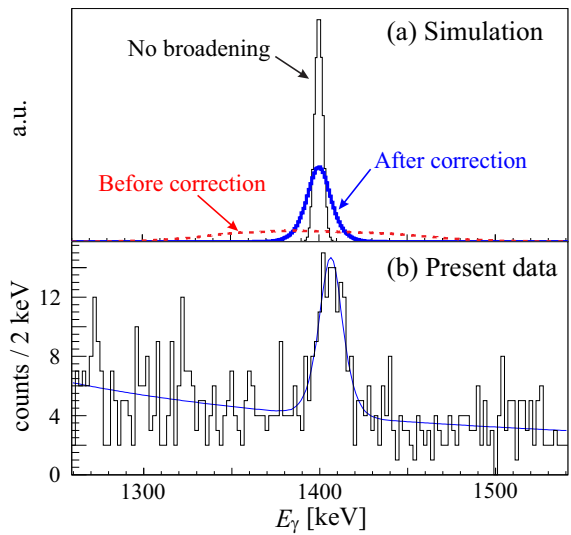}
\caption{
(color online).
(a) Simulated shapes of a 1.4 MeV $\gamma $-ray peak:
the thin black line corresponds to a $\gamma $ ray emitted at rest,
the dotted red line to a $\gamma $ ray emitted by
the recoiling $^4_\Lambda $He.
The thick blue line is the result of the Doppler-shift correction applied to the dotted one.
(b) shows the fit of the simulated peak shape
to the present data based on the simulated peak shape.
}
\label{He_Fit}
\end{center}
\end{figure}

Figure~\ref{He_Gamma} shows mass-gated $\gamma $-ray energy spectra.
Figure~\ref{He_Gamma}(a) and (b) are the spectra
without and with the Doppler shift correction, respectively,
when the highly unbound region ($E_{\rm ex} >$ +20 MeV) of $^4_\Lambda $He is selected.
Figure~\ref{He_Gamma}(c) is the spectrum without the Doppler-shift correction
for the $^4_\Lambda $He bound region.
Only after the event-by-event Doppler-shift correction,
the 1406-keV peak is clearly visible  as shown in Fig.~\ref{He_Gamma}(d).
The peak at 1406 keV is assigned
to the spin-flip $M1$ transition between the spin-doublet states,
$^4_{\Lambda }$He$(1^+ \to 0^+)$, 
because no other state which emits $\gamma $ rays is expected to be populated
in the selected excitation energy region.
This assignment is also consistent with the fact that the peak appears after
the Doppler shift correction.

Figure~\ref{He_Fit}(a) shows simulated $\gamma $-ray peak shapes.
The thin black line is for a $\gamma $ ray emitted at rest,
the dotted red line for a $\gamma $ ray emitted
immediately after the reaction where $^4_\Lambda $He has a maximum $\beta $
before slowing down in the target material,
and the thick blue line for a $\gamma$ ray with Doppler-shift correction applied to the dotted red line.
The observed peak shape shown in Fig.~\ref{He_Fit}(b) agrees with a simulated one
to which Doppler correction was applied
reflecting ambiguities
in the reconstructed vertex point and in the Ge detector positions.
The peak fitting result for the Doppler-shift-corrected spectrum
is presented in Fig.~\ref{He_Fit}(b).
The $\gamma $-ray energy and yield were obtained to be 
1406 $\pm$ 2(stat.) $\pm$ 2(syst.) keV and 95 $\pm$ 13 counts, respectively.
A dominant source of the systematic error comes from 
position inaccuracy of the reaction vertex and of the Ge detectors
for correcting the Doppler shift.  
The obtained yield is consistent with an expected value
based on a DWIA calculation of Ref.~\cite{Harada_private} within a factor of 3. 

% ** disscussion ** (this title will be removed) \\

In the present work, the $\gamma$-ray transition of 
$^4_\Lambda$He($1^+ \to 0^+$) was unambiguously observed,
and the excitation energy of $^4_\Lambda $He($1^+$) state
was precisely determined to be 1.406 $\pm$ 0.002 $\pm$ 0.002 MeV,
by adding a nuclear recoil correction of 0.2 keV.
%where the central energy was shifted by 0.2 keV from the measured $\gamma $-ray energy
%due to recoil of hypernucleus.
By comparing it to the previously measured spacing of 
$^4_\Lambda$H (1.09 $\pm 0.02$ MeV), the existence of CSB in $\Lambda N$ 
interaction has been definitively confirmed.
It is to be mentioned that two old experiments using stopped $K^-$
on $^6$Li and $^7$Li targets had reported hints of unassigned $\gamma$-ray peaks
at 1.42 $\pm$ 0.02 MeV \cite{BAMBERGER} and 1.45 $\pm$ 0.05 MeV \cite{H_1}, respectively.
It is presumed that those $\gamma$ rays came from $^4_\Lambda$He
produced as a hyperfragment.
By combining the emulsion data of $B_\Lambda$($^4_\Lambda$He$(0^+)$),
the present result gives 
$B_\Lambda$($^4_\Lambda$He($1^+$))= 0.98 $\pm$ 0.03 MeV
as shown in Fig.~\ref{He_level}.
By comparing it to $B_\Lambda$($^4_\Lambda$H($1^+$)) = 0.95 $\pm$ 0.04 MeV
obtained from the emulsion data of $B_\Lambda$($^4_\Lambda$H($0^+$))
and the $^4_\Lambda$H $\gamma $-ray data,
the present result leads to 
$\Delta B_\Lambda(1^+)$ = $B_\Lambda$($^4_\Lambda$He($1^+$)) $- B_\Lambda$($^4_\Lambda$H($1^+$))
= 0.03 $\pm$ 0.05 MeV.
% By combining the emulsion data of $B_\Lambda(0^+)$,
% the present result gives  
% $B_\Lambda$($^4_\Lambda$He($1^+$))= 0.98 $\pm$ 0.03 MeV.
% Considering also $B_\Lambda$($^4_\Lambda$H($0^+$))
% and the mean value of the $^4_\Lambda$H($1^+,0^+$) energy spacing,
% it is possible to obtain $B_\Lambda$($^4_\Lambda$H($1^+$)) = 0.95 $\pm$ 0.04 MeV,
% leading to  
Therefore, the CSB effect is strongly spin dependent,
being by one order of magnitude smaller in the $1^+$ state than in the $0^+$ state.
This demonstrates that the underlying $\Lambda$$N$ CSB
interaction has spin dependence.
Our finding suggests that $\Sigma$ mixing in $\Lambda$ hypernuclei
is responsible for the CSB effect since
the $1^+$ state in $^4_\Lambda$H/$^4_\Lambda$He receives
by one order of magnitude smaller an energy shift due to $\Lambda$-$\Sigma$ mixing than the $0^+$ state 
\cite{AKAISHI,HIYAMA}, which is caused by strong $\Lambda$$N$-$\Sigma$$N$ interaction in 
the two-body spin-triplet channel.

Recently, Gal \cite{Gal} estimated the CSB effect using a central-force
$\Lambda$$N$-$\Sigma$$N$ interaction
(D2 potential in Ref.~\cite{AKAISHI}),
in contrast to the widely-used tensor-force dominated
$\Lambda$$N$-$\Sigma$$N$ interaction in the Nijmegen OBE models.
Obtained $\Delta B_\Lambda(1^+)$ values are
%($\Delta B_\Lambda (0^+) \sim 250$ keV and
%$\Delta B_\Lambda (1^+) \sim 35$ keV)
in agreement with the present observation.
Further theoretical studies may reveal
not only the origin of the CSB effect
but also the properties of $\Lambda$-$\Sigma$ mixing in hypernuclei.

In summary, the J-PARC E13-1\textsuperscript{st} experiment clearly identified
a $\gamma$-ray transition from $^4_\Lambda$He
produced by the $^4$He$(K^-,\pi^-)$ reaction
and determined the energy spacing
between the spin-doublet states ($1^+,0^+$)
to be 1406 $\pm$ 2 (stat.) $\pm$ 2 (syst.) keV.
The apparent difference from the $^4_\Lambda$H spacing of 1.09 $\pm$ 0.02 MeV
and thus the existence of CSB in $\Lambda$$N$ interaction
have been confirmed only via the $\gamma$-ray measurement.
Combined with the emulsion data of $B_\Lambda(0^+)$,
the present result indicates a large spin dependence in the CSB effect,
by one order of magnitude larger in the $0^+$ state energy than in the $1^+$ state energy,
providing crucial information toward understanding
$\Lambda$$N$-$\Sigma$$N$ interaction and 
eventually baryon-baryon interactions.

%** Acknowledgment ** (this title will be removed) \\

We acknowledge experimental support from the J-PARC accelerator
and hadron experimental facility staff.
We thank SEIKO EG\&G and Fuji electric Co.~Ltd.~for support of our Ge detector system.
We thank Prof.~T.~Harada for theoretical inputs in designing the experiment.
This work is partially supported by Grant-in-Aid
Nos.~17070001, 21684011, 23244043, 24105003 and 24740138
for Scientific Research from the Ministry of Education Japan,
and a Grant-in-Aid No.~22$\cdot $3038 for JSPS Fellows,
and Basic Research (Young Researcher) No.~2010-0004752
from the National Research Foundation in Korea.
We acknowledge support from National Research Foundation,
WCU program of the Ministry of Education, Science and Technology (Korea),
and Center for Korean J-PARC Users (Grant No. 2013K1A3A7A060565).
we also thank KEKCC and SINET4.
\bibliography{PRL_He2015_0712}

\end{document}